\documentstyle[prb,aps,epsfig]{revtex}
\title{$\nu=2$ Bilayer Quantum Hall System in Tilted Magnetic Field}
\author{A.A. Burkov and A.H. MacDonald}
\address{\small Department of Physics, Indiana University, Bloomington, 
IN 47405}
\address{\small Department of Physics, The University of Texas at Austin, 
Austin, TX 78712} 
\date{\today}
\begin{document}
\maketitle
\begin{abstract}

We report on a theoretical study of $\nu=2$ bilayer quantum Hall systems
with a magnetic field that has a component parallel to the layers. 
As in the $\nu=1$ case, interlayer phase coherence is closely coupled to 
electron correlations and the Aharonov-Bohm phases introduced by a 
parallel magnetic field can have a strong influence on the ground state of 
the system.  
We find that response of a $\nu=2$ system to a 
parallel field is more subtle than that of a $\nu=1$ system because of 
the interplay between spin and layer degrees of freedom.
There is no commensurate-incommensurate transition as the parallel field
is increased.  Instead, we find a new phase transition which can occur 
in fixed parallel field as the interlayer bias potential is varied.
The transition is driven by the competition between canted 
antiferromagnetic order and interlayer phase coherence in the presence
of the parallel field.  We predict a strong singularity in the differential
capacitance of the bilayer which can be used to detect the phase transition.

\end{abstract}

\newpage

\section{Introduction}

There has recently been a great deal of theoretical work on
broken symmetry ground states in bilayer quantum Hall 
systems.~\cite{Fertig89,MacDonald90,Brey90,Wen92,Iwazaki,Moon95,Yang96,Zheng97,Dassarma98,Ezawa,Dassarma99,MacDonald99,Paredes99,Chang99,Schliemann00,Schliemann01,Rajaraman00,Radzihovsky01,Ramin01,Yogesh01,josephson} 
The simplest and most studied case has Landau level filling factor $\nu=1$.
For small enough layer separation these systems can have spontaneous interlayer phase coherence,
{\it i.e.} phase coherence in the absence of inter-layer tunneling.
This broken symmetry is driven by the improved inter-layer electronic correlations that it yields.  
Experimentally the existence of spontaneous interlayer phase coherence
in $\nu=1$ bilayers has been quite directly established in a series of recent experiments by 
Eisenstein and 
collaborators.~\cite{Eisenstein92,Murphy94,Eisenstein00,Eisenstein01,Kellogg01} 
Among the interesting phenomena
that have been associated with spontaneous interlayer phase coherence is a strong 
response to the Aharonov-Bohm phases produced when the magnetic field is tilted
from the normal to the layers and magnetic flux penetrates the space between the layers.~\cite{Moon95,Yang96,Murphy94,Read95,Hanna01,Papa02}
An in-plane field component alters the charge gap of the broken symmetry $\nu=1$ state,
and eventually leads to a commensurate-incommensurate phase transition which 
introduces solitons in the inter-layer phase field.
In this paper we present a theoretical analysis of the corresponding effects 
at Landau level filling factor $\nu=2$.

For perpendicular fields,
the $\nu=2$ bilayer's phase diagram is richer than at $\nu=1$ because both 
the pseudospin, used to descibe the {\em which layer} degree of freedom, and 
the real spin are important.  
The ground state has both spin and interlayer phase coherence 
broken symmetries~\cite{MacDonald99} and a very complex dependence on 
interlayer tunneling, bias potential, and Zeeman coupling external fields.
For finite tunneling the ground state can be described as a canted antiferromagnet,~\cite{Zheng97,Dassarma98,Dassarma99} in which spins
in opposite layers have opposing tilts away from the Zeeman field direction 
that are controlled by a competition between intralayer correlations, which 
favor ferromagnetic order within each layer, Zeeman coupling, which favors 
spin polarization along the magnetic field, and tunneling, which favors
opposite orientations of the spins in opposite layers.
The broken symmetry of this state has another aspect, however, 
which has usually been ignored in the literature---spontaneous phase coherence
between up(down)-spin electrons in one layer and down(up)-spin electrons 
in the other layer.  This order is spontaneous even at nonzero tunneling, 
unlike interlayer phase coherence in the $\nu=1$ case.  
The field that is conjugate to this order 
parameter, a spin-dependent tunneling field, is extremely weak in practice, 
removing a number of potentially interesting phenomena from experimental 
accessibility.  
This aspect of the broken symmetry is, however, key to understanding the subtle
response of a $\nu=2$ bilayer to an in-plane field that we address in this 
paper.
Experimental studies of $\nu=2$ bilayers have not yet produced experimental 
signatures 
of order or of phase transitions that are as stark as in the $\nu=1$ case. 
There are signatures of possible broken symmetry states in inelastic
light scattering~\cite{Pellegrini97} and transport~\cite{Sawada98,Khrapai00}
experiments, but there is so far no direct proof that the antiferromagnetic
and interlayer phase coherent orders do exist.  This study of in-plane field 
response is motivated by the expectation that signatures of the 
phase coherent aspect of the $\nu=2$ state order should exist. 
We find that the commensurate-incommensurate transition with increasing
parallel field that occurs in the $\nu=1$ case does not occur at $\nu=2$,
essentially because the $\nu=2$ phase coherence is off-diagonal in spin 
indices.
However we do find a new phase transition which can occur 
in fixed parallel field as the interlayer bias potential is varied,
which is a signature of the ground state broken symmetry. 
We predict a strong singularity in the differential
capacitance of the bilayer which can be used to detect this phase transition.

To understand the response of the $\nu=2$ bilayer to a parallel field, it is 
helpful to compare it with the corresponding response at $\nu=1$.
In order to enclose the magnetic flux produced by the in-plane field component,
bilayer system electrons must tunnel between layers.  
Because of macroscopic phase coherence, the properties of the ground state
at $\nu=1$ are extremely sensitive to weak parallel magnetic fields 
$B_{||}$ oriented in the plane of the bilayer even when the amplitude for 
tunneling is very small, as shown both 
theoretically~\cite{Yang96} and experimentally.~\cite{Murphy94} 
At very small parallel fields the ground state is commensurate,
that is the relative phase between the electrons 
in opposite layers develops a uniform gradient that follows
the gradient of the Aharanov-Bohm (AB) phase that multiplies the tunneling
matrix element in one convenient gauge choice.~\cite{Yang96} 
This gradient is oriented perpendicular 
to the in-plane field component, {\it i.e.} in the direction in which the 
AB phase accumulates linearly.  
In the commensurate state, the system preserves tunneling energy
at the expense of the interlayer exchange-correlation energy. 
At a certain critical value of the parallel field $B_{||}^*$ the
cost in exchange-correlation energy becomes too large and a phase transition to a soliton-lattice state occurs.  
At large $B_{||}^*$, the soliton-lattice state 
asymptotically approaches an incommensurate state which fully gives up the 
tunneling energy in order to preserve the inter-layer exchange-correlation 
energy. 
The phase winding length $L_{||}^*$ at the commensurate-incommensurate phase
transition is many times larger then the magnetic length $\ell$, 
indicative of the highly collective nature of this phenomenon. 

The more complex behavior we find at $\nu=2$ reflects the presence of 
both antiferromagnetic and interlayer phase coherence aspects, 
that are influenced by the parallel field in a different manner.
Indeed we find that the behavior of $\nu=2$ bilayer in a parallel field
is mostly determined by the competition between antiferromagnetism and 
interlayer phase coherence.  We find that for $\nu=2$ the commensurate state 
always has a lower energy than the incommensurate state and conclude that 
no soliton-lattice states of any type occur.  
The essential difference in the $\nu=2$ case is that 
the system has more freedom to adjust to the parallel field than in  
the $\nu=1$ case, and can preserve a large fraction of its tunneling energy in
the commensurate state even as $B_{||}\rightarrow\infty$. 
The driving force for the new transition we predict, which we expect to be of 
first order, is the competition between 
canted antiferromagnetic order and interlayer phase coherence in the parallel magnetic field. 
The transition results in discontinuities in the order parameters
and a large singularity in the interlayer differential capacitance $C_{int}$ 
of the system.  Since $C_{int}$ is a relatively easily measurable quantity, we expect it to be 
possible to observe this phase transition experimentally.
Observation of this phase transition would be the first direct experimental proof of the existence of
the canted antiferromagnetic and spontaneous interlayer phase coherent
ordering in $\nu=2$ bilayers. 
It is interesting to note that similar parallel-field-induced first order 
transition manifested by a diverging differential capacitance, was recently 
predicted in $\nu=1$ bilayers.~\cite{Radzihovsky01,Ramin01} 
Despite the similarity however, the mechanism of the transition is very 
different in our case.  

Our paper is organized as follows.  In Section~\ref{s2} we introduce the 
class of 
unrestricted Hartree-Fock variational wavefunctions we consider.  
Our calculations become exact if a classical approximation is used for spin and 
pseudospin variables, charge fluctuations in the incompressible
$\nu=2$ state are neglected, and broken translational symmetry, 
which would not be anticipated in this case unless there is a 
commensurate-incommensurate 
phase transition, can be ruled out as a possibility.
The variational wavefunctions are specified by two arbitrary 
4-component spinor wavefunctions, as in the approach used by one
of us~\cite{MacDonald99} in the absence of an in-plane field.  Here, however, 
we allow each of the variational parameter phases to have uniform gradients of
arbitrary magnitude.  
Our conclusions are based on the minimization of the corresponding 
energy functional, which leads to a set of Hartree-Fock single-particle 
equations, that are derived in Section~\ref{s2}.
Section~\ref{s3} reports results of the numerical solution of the HF 
equations to locate 
minima of the Hartree-Fock energy functional.  
The behavior of order parameters as a function of in-plane field and 
physically adjustable
external fields, the inter-layer bias potential in particular, is discussed.
We focus here on the differential bilayer capacitance and on the anomaly it
shows at the system's first order phase transition.
We conclude in Section~\ref{s4} with a short resume of our results.

\section{Unrestricted Hartree-Fock theory of the $\nu=2$ bilayer}    
\label{s2}
The physics of the broken symmetry states at integer filling factors in quantum
Hall systems is simplified by the fact that they are incompressible states
that have a gap for charged excitations.   We take advantage of this property
by using a Hartree-Fock approximation that 
neglects charge fluctuations completely and amounts to using a    
classical approximation for the remaining spin and pseudospin degrees of 
freedom.  
We assume that only two single-particle states are relevant in the growth 
direction of the bilayer, one localized in each well, so that we can use a 
pseudospin to represent this translational degree of freedom.
Electrons in $\nu=2$ bilayer system can then be described as being in a 
coherent superposition of spin and pseudospin up and down eigenstates. 
Our variational HF wavefunction has the form~\cite{MacDonald99}
\begin{equation}
\label{1}
|\Psi[z]\rangle\,=\,\prod_{i,X}\,\left(\,\sum_{k=1,4}\,
z^i_{kX}c^{\dagger}_{kX}\,\right)\,| 0 \rangle.
\end{equation}
Here $X$ is a Landau gauge Lowest Landau Level (LLL) orbital index and
k is a spin-pseudospin state label 
($k=1$ is a spin-up electron in the top layer, $k=2$ a spin-down electron
in the top layer, $k=3$ a spin-up in the bottom layer, and $k=4$ a spin-down
in the bottom layer). 
The spin quantization axis is along the magnetic field direction.
The index $i=1,2$ labels the two lowest energy eigenstates of the HF 
Hamiltonian which we derive below. 

We allow the coefficients $z^i_{kX}$ to have the following 
dependence on the LLL orbital index
\begin{equation}
\label{2}
z^i_{kX}\,=\,z^i_k\,e^{iQ_kX}.
\end{equation}
This choice generates translationally invariant spin and pseudospin spiral 
states and excludes the possibility of non translationally invariant states 
containing spin and pseudospin vortices or solitons.
This restriction will be justified {\em post factum} by the fact that there 
is no commensurate-incommensurate phase transition in our system. 

The microscopic Hamiltonian for lowest Landau level electrons in bilayers 
has the following form, 
\begin{eqnarray}
\label{3}
H\,&=&\,\sum_{k_1,k_2,X}\,c^{\dagger}_{k_1 X} h^0_{k_1 k_2}
c_{k_2X}\,\nonumber \\
&+&\,\frac{1}{2}\, \sum_{k_1,k_2}\,\sum_{X_1,X_2,X'_1,X'_2}\,
c^{\dagger}_{k_1 X_1} c^{\dagger}_{k_2 X_2}
c_{k_2 X'_2} c_{k_1 X'_1}\,\langle k_1 X_1,k_2 X_2 | V |
k_1 X'_1,k_2 X'_2 \rangle.
\end{eqnarray}
Here $V$ is the 2D Coulomb interaction which is different if electrons 
are in the same or different layers and $h^0$ is the single-particle
part of the Hamiltonian which is given by
\begin{equation}
\label{4}
h^0\,=\,-(\Delta_V/2)\tau^z\,-\,(\Delta_t/2)\tilde \tau^x\,-\,(\Delta_z/2)
\sigma^z.
\end{equation} 
Here $\Delta_V$, $\Delta_t$ and $\Delta_z$ are the interlayer bias, 
the tunneling amplitude (single-particle symmetric-antisymmetric gap) and the
Zeeman splitting respectively. 
We assume that the interlayer tunneling amplitude is always nonzero.
Unlike the $\nu=1$ case, where there is a spontaneous interlayer phase
coherence at zero tunneling, broken symmetry states in $\nu=2$ case
occur over a range of $\Delta_t$ values.
$\sigma$ and $\tau$ are $4\times4$ spin and pseudospin Pauli matrices.
In a parallel magnetic field, $B_{||}$, the tunneling matrix elements acquire
an additional phase factor~\cite{Yang96} 
$e^{\pm iQ_{||}X}$, where $Q_{||}=B_{||}d/B_{\perp} \ell^2$,
and $d$ is the interlayer distance.  
It is easy to verify that these phase factors
incorporate the AB phases associated with closed paths that enclose flux
produced by the in-plane field.  
They are incorporated in the Hamiltonian by replacing
the $\tau_x$ pseudospin Pauli matrix by  
\begin{equation}
\label{5a}
\tilde \tau^x=\left(
\begin{array}{cccc}
0 & 0 & e^{i Q_{||}X} & 0\\
0 & 0 & 0 & e^{iQ_{||}X}\\
e^{-i Q_{||}X} & 0 & 0 & 0\\
0 & e^{-i Q_{||}X} & 0 & 0
\end{array}
\right).
\end{equation}
We assume that as the sample is tilted, the perpendicular component
of the magnetic field is kept constant, since we are interested in phenomena 
that occur at the fixed Landau level filling factor $\nu=2$.  
The Zeeman coupling constant therefore depends on the parallel component 
of the magnetic field as
\begin{equation}
\label{4.1} 
\Delta_z=\Delta_z^0\sqrt{1+\left(\frac{B_{||}}{B_{\perp}}\right)^2}.
\end{equation}

At a particular value of the in-plane field, and external coupling parameters,
we determine the ground state by calculating the expectation 
expectation value of the Hamiltonian (\ref{3}) in the 
many-body state (\ref{1}) and optimizing it with respect to the variational 
parameters $z^i_{k}$ and $Q_k$. 
It is important to realize that unlike the case with no 
parallel field,~\cite{MacDonald99}
the variational parameters cannot be assumed to be real.
In the present case the Hartree-Fock energy must be optimized with respect 
to {\it both absolute values $z^i_k$ and phases $Q_kX$ of all of the 
amplitudes $z^i_{kX}$}.  
The Hartree-Fock energy is given by
\begin{eqnarray}
\label{5}
E\,&=&\,-\frac{1}{2}\sum_{k_1,k_2,X}\,\left\{\Delta_V\tau^z_{k_1k_2}
+\Delta_t\tilde \tau^x_{k_1k_2}+\Delta_z\sigma^z_{k_1k_2}
\right.\nonumber \\
&-&\left. \,H\tau^z_{k_1k_2}[Tr(\rho\tau^z)]\,+\,
\rho_{k_1k_2}e^{i(Q_{k_1}-Q_{k_2})X} F_{k_1k_2}(Q_{k_1}-Q_{k_2})\right\}
\rho_{k_2k_1}e^{i(Q_{k_2}-Q_{k_1})X}.
\end{eqnarray}
Here $\rho$ is the Hartree-Fock density matrix
\begin{equation}
\label{6}
\rho_{k_1k_2}\,=\,\sum_{i=1}^2\,z^i_{k_1} z^i_{k_2},
\end{equation}
$H$ is the parameter characterizing the Hartree (electrostatic)  energy 
\begin{equation}
\label{7}
H\,=\,(2\pi l^2)^{-1}V_-({\bf q}=0),
\end{equation}
where $V_-=(V_S-V_D)/2$ and $V_{S,D}$ are the Coulomb interactions between
electrons in the same or different layers. 
For strictly 2D layers $V_{S,D}$ are given by
\begin{eqnarray}
\label{8}
V_S({\bf q})\,&=&\,\frac{2\pi e^2}{\epsilon q} \nonumber \\
V_D({\bf q})\,&=&\,\frac{2\pi e^2}{\epsilon q}e^{-qd}.
\end{eqnarray}
Parameters $F_{kk'}$ characterizing the exchange matrix elements of the Hamiltonian are given by 
\begin{equation}
\label{9}
F_{k_1k_2}(Q_{k_1}-Q_{k_2})\,=\,\int \frac{d^2q}{(2\pi)^2}V_{k_1k_2}({\bf q})
e^{-q^2l^2/2-
i(Q_{k_1}-Q_{k_2})q_yl^2},
\end{equation}
where $V_{k_1k_2}$ is equal to $V_S$ when the labels refer to the same layer 
and to $V_D$ when the labels refer to different layers.  
Eq.~(\ref{9}) captures the 
reduction in exchange energy that occurs when different components of the 
variational spinors have different wavevectors.
Note that only the tunneling part of the Hartree-Fock energy depends on the 
LLL orbital label $X$. 

Minimizing the HF energy with respect to $z^i_{kX}$ one obtains the following
Hartree-Fock single-particle Hamiltonian
\begin{equation}
\label{10}
h^{HF}_{k_1k_2}(X)\,=\,h^0_{k_1k_2}(X)\,+\,H\tau^z_{k_1k_2}[Tr(\rho\tau^z)]
\,-\,\rho_{k_1k_2}e^{i(Q_{k_1}-Q_{k_2})X} F_{k_1k_2}(Q_{k_1}-Q_{k_2}).
\end{equation}
The most convenient strategy for numerical calculations is to solve the 
Hartree-Fock equations 
\begin{equation}
\label{11}
\sum_{k_2}h^{HF}_{k_1k_2}(X) z^i_{k_2} e^{i(Q_{k_2}-Q_{k_1})X}\,=\,
\epsilon^i z^i_{k_1}
\end{equation}
to find extrema of the energy functional for given values of the 
$Q_k$, and then optimize the $Q_k$ values. 
 
Three classes of solutions of (\ref{11}) exist.  In each case,
as implied by the notation of Eq.~(\ref{11}), the single-particle 
eigenvalues are independent of $X$. 
\begin{enumerate}
\item Fully commensurate solutions, that fully preserve the tunneling energy 
for both spin directions.  
In this case $Q_1-Q_3=Q_2-Q_4=Q_{||}$ to capture the tunneling energy. 
It will be critical below that $Q_1-Q_2$ and $Q_3-Q_4$ can still be varied 
arbitrarily, at a cost in 
exchange energy within each well but without any cost in tunneling energy. 
\item Fully incommensurate solutions in which all phase gradients are set to 
zero.
Strictly speaking these solutions solve the Hartree-Fock equations only if the 
tunneling amplitude is set to zero.  
From a variational point of view, these solutions 
may be regarded as approximations to the soliton lattice states that could 
occur in the system, for which the tunneling contribution to the energy 
vanishes when $Q_{||}$ is much larger than the critical value at which the 
commensurate-incommensurate transition occurs.  
We find below that, unlike the $\nu=1$ case, incommensurate 
solutions always have higher energy than commensurate solutions and 
conclude on this
basis that there is no commensurate-incommensurate phase transition 
for $\nu=2$.
\item Partially commensurate/incommensurate solutions with tunneling 
energy preserved for only one of the spin directions.
Here we have $Q_1-Q_3=0,\, Q_2-Q_4=Q_{||}$ or
$Q_1-Q_3=Q_{||},\,Q_2-Q_4=0$.
\end{enumerate}  
Our numerical calculations demonstrate that
the fully commensurate solution is always the ground 
state.  The main reason that the commensurate-incommensurate transition does 
not happen in our case is
that at $\nu=2$ an additional degree of freedom---the phase difference
between spin-up and -down electrons in the same layer is available,
allowing the system to keep the tunneling energy without suffering
almost any loss of exchange energy.  
In what follows we will discuss only commensurate solutions.  

\section{Results and Discussion}
\label{s3}
We have solved the HF equations (\ref{11}) numerically for different values of
parallel magnetic field, interlayer bias potential, and tunneling amplitude.
We keep the Zeeman splitting at zero parallel field fixed at 
$\Delta_z^0=0.01$ in units of $e^2/\epsilon \ell$ 
since it is difficult to deviate far from this value in experimental systems.
We keep the interlayer distance $d$ equal to the magnetic length, since it
is also difficult to vary this parameter widely. 
Since the same-spin interlayer phase difference is fixed by tunneling, there
is only one free phase gradient in commensurate state calculations, 
the gradient of the phase difference between the up- and down-spin electrons 
in the same layer, which we will denote by $Q$. 
All the phase difference gradients can be expressed
in terms of this gradient and the one due to the parallel field 
$Q_{||}=B_{||}d/B_{\perp} \ell^2$ as follows
\begin{eqnarray}
\label{12}
&&Q_1-Q_3=Q_2-Q_4=Q_{||} \nonumber \\ 
&&Q_1-Q_2=Q_3-Q_4=Q \nonumber \\
&&Q_1-Q_4=Q_{||}+Q \nonumber \\
&&Q_2-Q_3=Q_{||}-Q.
\end{eqnarray}
We see from Eq.~(\ref{12}) that the intralayer phase gradient $Q$   
and the interlayer phase gradient due to the parallel field $Q_{||}$ are 
coupled. 
The optimal value of $Q$ is determined mostly by
an interplay between the intralayer spin exchange energy, 
proportional to $F_S(Q)$, and the interlayer 
exchange energy, proportional to $F_D(Q_{||}+Q)$ and $F_D(Q_{||}-Q)$. 

\begin{figure}
\begin{center}
\epsfxsize=4in
\epsffile{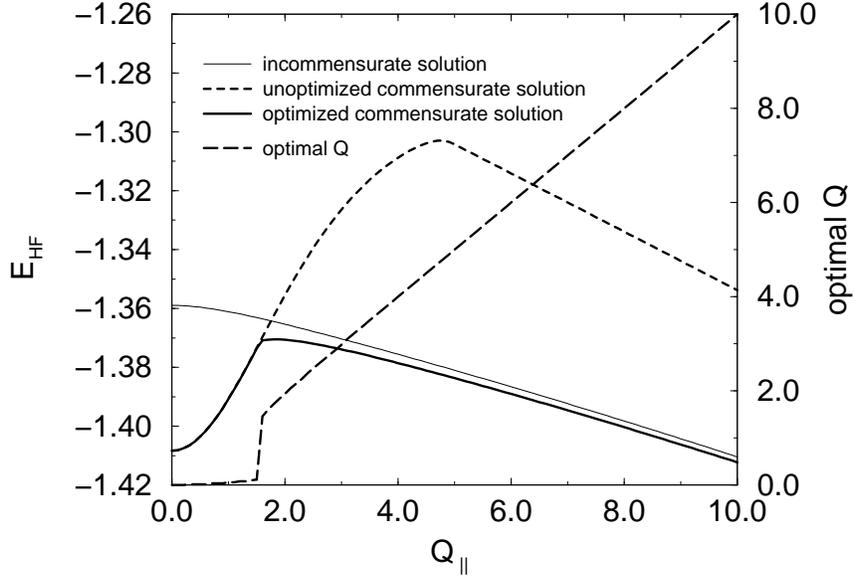}
\caption{Hartree-Fock energy of the commensurate state optimized 
with respect to the intralayer phase gradient $Q$ (thick solid line),
incommensurate state ($Q=0$) (thin solid line) and commensurate state at $Q=0$
(short-dashed line)
for $\Delta_t=0.1$ and $\Delta_V=1.0$. Also shown is the optimal value of 
the intralayer phase gradient $Q$ (long-dashed line). 
There is a commensurate-incommensurate 
transition if one keeps $Q$ equal to zero. However, when commensurate state 
is optimized with respect to $Q$ the transition is avoided.}
\label{fig1}
\end{center}
\end{figure}

\begin{figure}
\begin{center}
\epsfxsize=4in
\epsffile{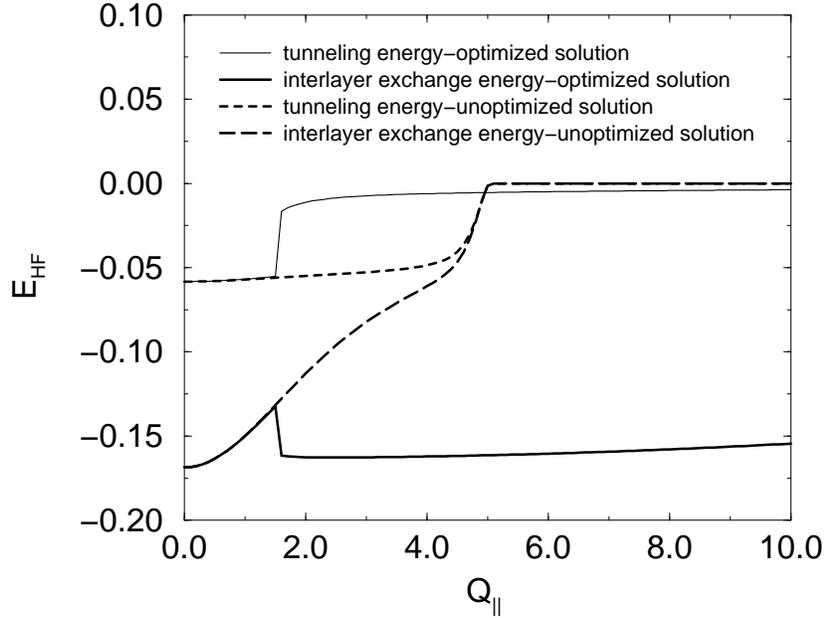}
\caption{Interlayer exchange HF energy (thick solid line) and tunneling energy
(thin solid line) for the fully optimized commensurate solution and 
the corresponding energies for the unoptimized solution ($Q=0$) (dashed lines)
at $\Delta_t=0.1$, $\Delta_V=1.0$. Fully optimized commensurate state gains 
interlayer exchange energy at the same time preserving nonzero tunneling 
energy.}
\label{fig2}
\end{center}
\end{figure}

At each value of $B_{||}, \Delta_t, \Delta_V$ and $Q$ we find the 
self-consistent solution of the HF equations and optimize it with respect 
to $Q$.   
In Fig.~\ref{fig1} we plot the total Hartree-Fock energy as a function of 
$Q_{||}$ for $\Delta_t=0.1$ and $\Delta_V=1.0$, comparing it to the 
Hartree-Fock  
energy of the incommensurate state, {\it i.e.} the energy in the absence of 
tunneling. 
If the commensurate state energy crossed above the incommensurate state energy
at some value of $Q_{||}$, we would expect a commensurate-incommensurate phase transition to occur.  
Indeed, we see in Fig.~\ref{fig1} that this is exactly what does occur when
$Q$ is fixed at zero for each value of $Q_{||}$.  
However, when $Q$ is properly 
optimized at each value of $Q_{||}$, it becomes clear that the transition is 
circumvented. 
The cusp in the dependence of the HF energy of the commensurate state 
on $Q_{||}$ is a signature of a first order transition as is evident 
from the plot of the optimal value of $Q$ vs. $Q_{||}$ in the same figure.
By going from a state with $Q \sim 0$ to $Q \sim Q_{||}$ the system gains
spin-off-diagonal interlayer exchange energy ($F_D(Q_{||}-Q)$) without losing 
all of its tunneling energy. The new state is very close to the 
incommensurate state, but has slightly lower energy, since tunneling energy 
is small but still nonzero. This is illustrated in Fig.~\ref{fig2} where
interlayer exchange and tunneling contributions to the total HF energy 
are plotted for both optimized and unoptimized commensurate states. 
Thus the commensurate-incommensurate transition is avoided. 

Once we have obtained the optimal HF solution, various physical observables 
can be 
evaluated. In particular we are interested in the behavior of the canted 
antiferromagnetic
order parameter $O_{zx}=\langle \tau^z\sigma^x\rangle$, the interlayer 
phase coherence order parameter $O_{xx}=-\langle\tau^x\sigma^x\rangle$
and the interlayer differential capacitance $C_{int}=d\langle\tau^z\rangle/ d\Delta_V$. The latter quantity is experimentally accessible.

In the absence of an in-plane field the $\nu=2$ bilayer phase diagram
is already rich with a continuous phase transition occuring between broken
symmetry and normal ground states along a boundary that is sensitive to 
all external field parameters, particularly the interlayer bias potential.
We find that in a parallel magnetic field
there is in addition a first-order transition characterized by  
a discontinuous change in $Q$.
The canted antiferromagnet aspect of the ordered state is favored by  
the intralayer spin exchange interaction which is maximized at $Q=0$. 
At a zero parallel field the same is true for the interlayer exchange 
which favors the interlayer phase coherent aspect of the broken symmetry 
state's order. 
However, at a finite field the interlayer exchange energy
of the commensurate state is maximized at $Q=Q_{||}$.
A nontrivial optimal value of $Q$ exists, depending on the relative 
strength of the two order parameters, a competition that is tunable by the 
external interlayer bias potential.     

\begin{figure}
\begin{center}
\epsfxsize=4in
\epsffile{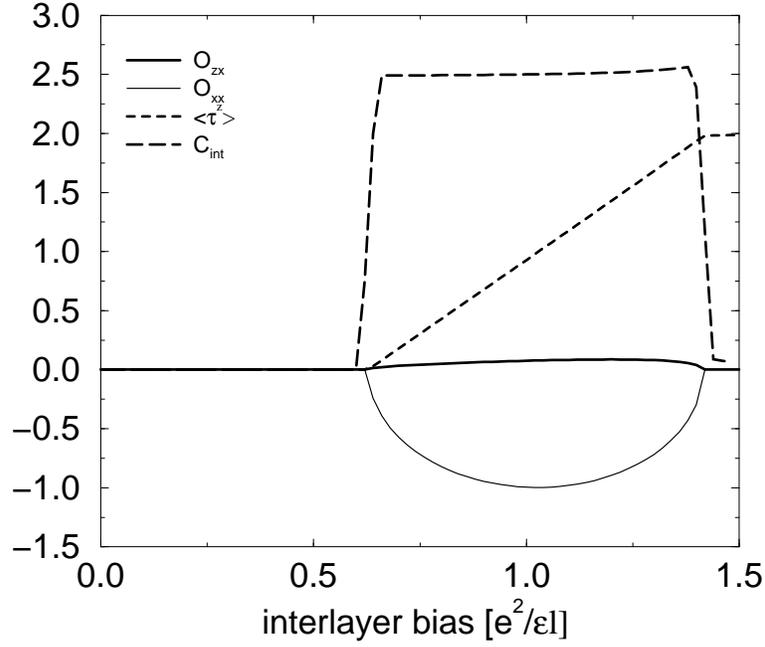}
\caption{Order parameters $O_{zx}$ (thick solid line), 
$O_{xx}$ (thin solid line), charge imbalance (dashed line) and differential
capacitance (long dashed line) for $\Delta_t=0.05$, and tilt angle 
$\Theta = 72.0 \deg$. There are two continuous order-disorder transitions,
where differential capacity has a discontinuity but no divergence. 
The canted antiferromagnetic order is too weak 
to compete with interlayer phase coherence, hence no first order transition
is observed.}
\label{fig3}
\end{center}
\end{figure}

\begin{figure}
\begin{center}
\epsfxsize=4in
\epsffile{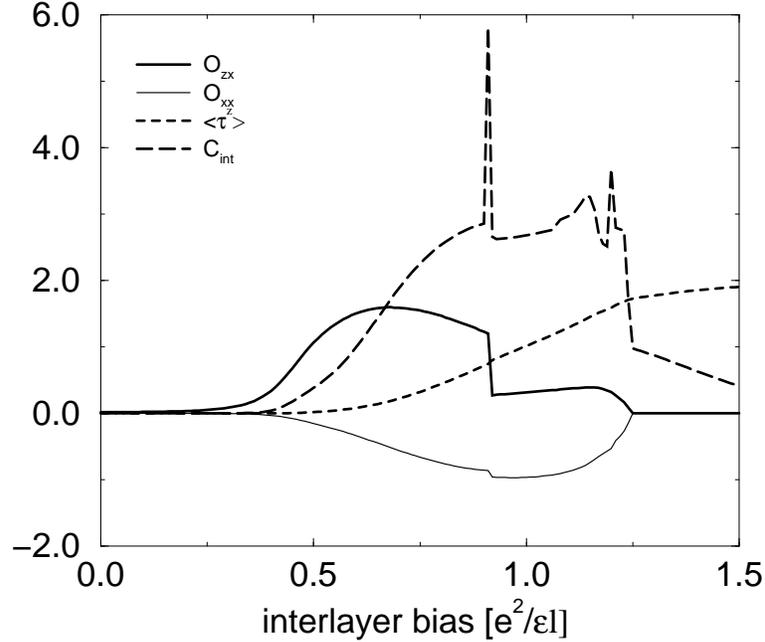}
\caption{Order parameters $O_{zx}$ (thick solid line), 
$O_{xx}$ (thin solid line), charge imbalance (dashed line) and differential
capacitance (long dashed line) for $\Delta_t=0.1$, and a tilt angle 
$\Theta = 58.0 \deg$. There are two first order transitions between a state
where canted antiferromagnetic order dominates to a state with interlayer 
phase coherence dominating and back. Transitions are manifested by divergences
in the interlayer differential capacitance.} 
\label{fig4}
\end{center}
\end{figure}

\begin{figure}
\begin{center}
\epsfxsize=4in
\epsffile{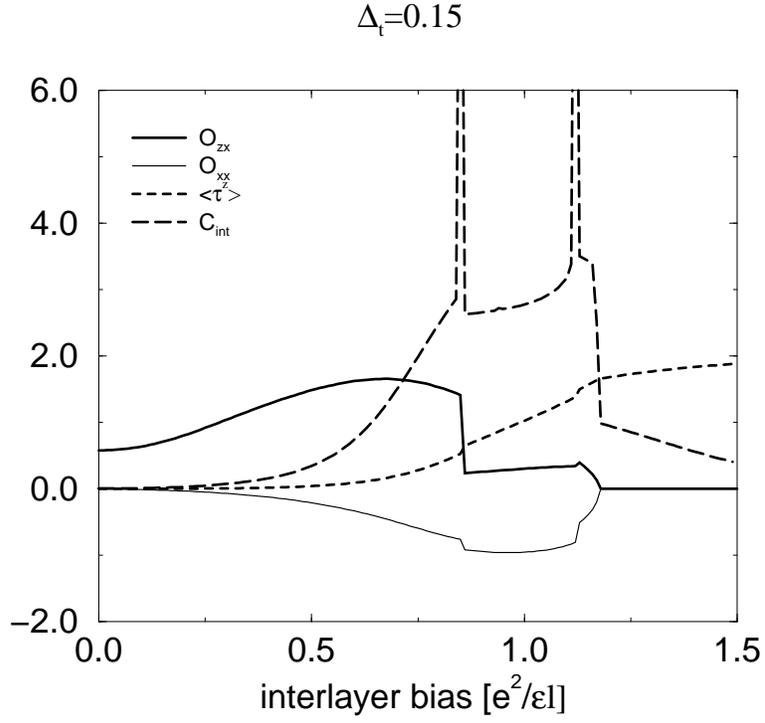}
\caption{Order parameters $O_{zx}$ (thick solid line), 
$O_{xx}$ (thin solid line), charge imbalance (dashed line) and differential
capacitance (long dashed line) for $\Delta_t=0.15$, and a 
tilt angle $\Theta = 66.5 \deg$. The first order phase transitions due to 
the competition between canted antiferromagnetism and interlayer phase 
coherence have become more pronounced.} 
\label{fig5}
\end{center}
\end{figure}

\begin{figure}
\begin{center}
\epsfxsize=4in
\epsffile{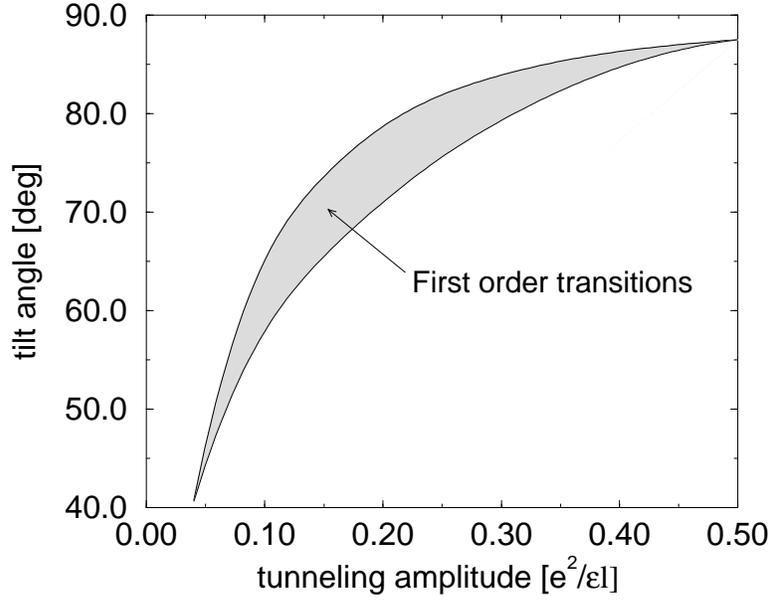}
\caption{First order phase transitions driven by the bias voltage are 
observed in the shaded region in the interlayer tunneling---tilt angle phase 
space.} 
\label{fig6}
\end{center}
\end{figure}

Our numerical results are shown in Figs.\ref{fig3}-\ref{fig6}.
Fig.\ref{fig3} illustrates the system's dependence on bias potential 
for a relatively small tunneling amplitude $\Delta_t = 0.05$ and a tilt angle 
$\Theta= \tan^{-1}(B_{||}/B_{\perp}) = 72.0\,\deg$, 
where we do not observe first order transitions.
As the bias potential is increased, there are two order-disorder 
transitions at which the differential capacitance has a discontinuity, but
no divergence.  In this case, the canted antiferromagnet order parameter
is very small and $Q=Q_{||}$ in the broken symmetry region,
since it allows for a greater gain in the interlayer exchange energy.  
In the disordered phase (where order parameters are zero) the HF energy 
does not depend on the intralayer
phase gradient $Q$, so that no singularity is observed at 
the order-disorder transition.
Fig.\ref{fig4} shows the same dependence at $\Delta_t=0.1$ for a  
tilt angle $\Theta = 58.0\, \deg$.
In this case there is a discontinuity in the charge transferred between
layers by the bias potential and a corresponding delta-function contribution 
to the differential capacitance.  This feature is associated with a 
shift in the value of $Q$ at which the global energy minimum occurs from 
a small value $Q \sim 0$ to $Q \sim Q_{||}$.
These two ground states have different equilibrium charge imbalances 
between the layers, hence the discontinous change in the charge imbalance at 
the transition.  
A similar feature occurs at a larger bias potential when
the global minimum shifts back to small $Q$.
At larger value of $\Delta_t$, as illustrated in Fig.\ref{fig5}, the 
two peaks have comparable strength.

First order transitions occur as a function of bias voltage 
in the shaded region in the tilt 
angle---$\Delta_t$ phase space in Fig.~\ref{fig6}.
Discontinuous transitions do not occur for  
very small tunneling amplitudes because the canted antiferromagnetic aspect of 
the order is relatively weak so that it is always preferable to 
have $Q \sim Q_{||}$ to optimize interlayer exchange energy.
For very strong tunneling the first order transitions occur 
only for tilt angles nearly equal $90\,\deg$, because the canted 
antiferromagnetic order dominates the interlayer phase coherence and
the cost in interlayer exchange needs to be very high to trigger the 
transition. 
The first order transition region has a high-tilt boundary since 
the canted antiferromagnetic order is weakened by tilting the sample 
due to the dependence of the Zeeman coupling on the parallel component 
of the field (see Eq.(\ref{4.1})).
Fig.~\ref{fig6} was obtained by sweeping the bias voltage at fixed values of
the tunneling amplitude 
and tilt angle and observing if the first order transitions and the 
corresponding differential capacitance singularities were present.
Since we have probed a limited number of points in the tunneling 
amplitude---tilt angle phase space, the curves in Fig.~\ref{fig6} are 
approximate. 
The corresponding error can be estimated to be $\sim 1\,\deg$. 

Since interlayer capacitance measurements are relatively straightforward, 
the observation of the above described divergences would be a very direct 
and unambiguous proof that the theoretically predicted broken symmetries
in $\nu=2$ bilayers do indeed exist.  

\section{Summary}
\label{s4}
The ordered ground state of $\nu=2$ bilayer quantum Hall systems 
can be regarded as a canted antiferromagnet or as a state with 
spontaneous coherence between states of opposite spin in opposite layers.
In the case of $\nu=1$ systems, competition between the 
tunneling energy and interlayer correlation energy 
in tilted magnetic fields leads to commensurate-incommensurate 
phase transition and a large reduction in the charge excitation gap.
In this paper, we have considered the behavior of the $\nu=2$ bilayer
quantum Hall system in a parallel magnetic field using an unrestricted
Hartree-Fock approximation.    
We have found that it differs strongly from the corresponding behavior of
$\nu=1$ bilayers.
The commensurate-incommensurate phase transition does not occur.
Due to the spin-off-diagonal nature of the interlayer phase coherence,
the cost in exchange energy may always be kept low enough for commensurate 
state to remain the ground state.   
We find that in a certain range of tunneling amplitudes first order transitions
can occur as a function of bias voltage at which intralayer 
correlations are improved and interlayer correlations are weakened and 
vice versa. 
The transitions are manifested by discontinuities in the interlayer bias 
dependence of the order parameters 
and are responsible for singularities that we predict 
in the interlayer differential capacitance.  
Observations of these phase transitions would provide a direct
verification of broken symmetry states in $\nu=2$ bilayer quantum
Hall ferromagnets.
As a final note, we would like to point out that the same first order 
transitions can be observed
at a fixed bias voltage by changing the tilt angle. 
We have intentionally limited our work to transitions driven by the bias 
voltage since experimentally it is much easier to change the bias at a fixed
tilt angle than vice versa.

This work was supported by the Welch Foundation, by the Indiana 21st 
Century Fund,
and by the National Science Foundation under grant DMR0115947.
AHM acknowledges a helpful conversation with Luis Brey.

\newpage



\newpage
\begin{thebibliography}{99}
\bibitem{Fertig89} H. A. Fertig, Phys. Rev. B {\bf 40}, 1087 (1989).
\bibitem{MacDonald90} A. H. MacDonald, P. M. Platzman, G. S. Boebinger,
Phys. Rev. Lett. {\bf 65}, 775 (1990).
\bibitem{Brey90} L. Brey, Phys. Rev. Lett. {\bf 65}, 903 (1990).
\bibitem{Wen92} X. G. Wen, A. Zee, Phys. Rev. Lett. {\bf 69}, 1811 (1992).
\bibitem{Iwazaki} Z.F. Ezawa, A. Iwazaki, Int. J. Mod. Phys. B {\bf 19},
3205 (1992); Z.F. Ezawa, A. Iwazaki, Phys. Rev. B {\bf 47}, 7295 (1993).
\bibitem{Moon95} K. Moon, H. Mori, Kun Yang, S.M. Girvin, A.H. MacDonald,
L. Zheng, D. Yoshioka, Shou-Cheng Zhang, Phys. Rev. B {\bf 51}, 5138 (1995).
\bibitem{Yang96} Kun Yang, K. Moon, Lotfi Belkhir, H. Mori, S.M. Girvin,
A. H. MacDonald, L. Zheng, D. Yoshioka, Phys. Rev. B {\bf 54}, 11644 (1996).
\bibitem{Zheng97} L. Zheng, R.J. Radtke, S. Das Sarma, Phys. Rev. Lett.
{\bf 78}, 2453 (1997).
\bibitem{Dassarma98} S. Das Sarma, S. Sachdev, L. Zheng, Phys. Rev. B {\bf 58},
4672 (1998).
\bibitem{Ezawa} Z.F. Ezawa, Phys. Rev. Lett. {\bf 82}, 3512 (1999); 
Phys. Lett. A {\bf 249}, 223 (1998).
\bibitem{Dassarma99} L. Brey, E. Demler, S. Das Sarma, Phys. Rev. Lett. 
{\bf 83}, 168 (1999).
\bibitem{MacDonald99} A.H. MacDonald, R. Rajaraman, T. Jungwirth,
Phys. Rev. B {\bf 60}, 8817 (1999).
\bibitem{Paredes99} B. Paredes, C. Tejedor, L. Brey, L. Martin-Moreno,
Phys. Rev. Lett. {\bf 83}, 2250 (2000).
\bibitem{Chang99} Min-Fong Yang, Ming-Che Chang, Phys. Rev. B {\bf 60}, 
R13985 (1999). 
\bibitem{Schliemann00} J. Schliemann, A.H. MacDonald, Phys. Rev. Lett.
{\bf 84}, 4437 (2000).
\bibitem{Schliemann01} J. Schliemann, S.M. Girvin, A.H. MacDonald,
Phys. Rev. Lett. {\bf 86}, 1849 (2001).
\bibitem{Rajaraman00} S. Ghosh, R. Rajaraman, Phys. Rev. B {\bf 63}, 
035304 (2000).
\bibitem{Radzihovsky01} L. Radzihovsky, Phys. Rev. Lett. {\bf 87}, 236802
(2001).
\bibitem{Ramin01} M. Abolfath, L. Radzihovsky, A.H. MacDonald, 
cond-mat/0110049 (2001).
\bibitem{Yogesh01} Y.N. Joglekar, A.H. MacDonald, Phys. Rev. B {\bf 64},
155315 (2001).
\bibitem{josephson} L. Balents, L. Radzihovsky, Phys. Rev. Lett. {\bf 86},
1825 (2001); A. Stern, S. M. Girvin, A. H. MacDonald, N. Ma, {\it ibid.} 
{\bf 86}, 1829 (2001); M. M. Fogler, F. Wilczek, {\it ibid.} {\bf 86}, 1833
(2001); Y. N. Joglekar, A. H. MacDonald, {\it ibid.} {\bf 87}, 196802
(2001). 
\bibitem{Burkov02} A.A. Burkov, A.H. MacDonald, cond-mat/0201355 (2002).
\bibitem{Eisenstein92} J.P. Eisenstein, G.S. Boebinger, L.N. Pfeiffer,
K.W. West, Song He, Phys. Rev. Lett. {\bf 68}, 1383 (1992).
\bibitem{Murphy94} S.Q. Murphy, J.P. Eisenstein, G.S. Boebinger, 
L.N. Pfeiffer, K.W. West, Phys. Rev. Lett. {\bf 72}, 728 (1994).
\bibitem{Eisenstein00} I.B. Spielman, J.P. Eisenstein, L.N. Pfeiffer,
K.W. West, Phys. Rev. Lett. {\bf 84}, 5808 (2000).
\bibitem{Eisenstein01} I.B. Spielman, J.P. Eisenstein, L.N. Pfeiffer, 
K.W. West, Phys. Rev. Lett. {\bf 87}, 0368031 (2001).
\bibitem{Kellogg01} M. Kellogg, I.B. Spielman, J.P. Eisenstein, L.N. Pfeiffer,
K.W. West, cond-mat/0108403 (2001).
\bibitem{Read95} N. Read, Phys. Rev. B {\bf 52}, 1926 (1995).
\bibitem{Hanna01} C.B. Hanna, A.H. MacDonald, S.M. Girvin, Phys. Rev. B
{\bf 63}, 125305 (2001).
\bibitem{Papa02} E. Papa, A.M. Tsvelik, cond-mat/0201343 (2002).
\bibitem{Pellegrini97} V. Pellegrini, A. Pinczuk, B.S. Dennis, A.S. Plaut,
L.N. Pfeiffer, K.W. West, Phys. Rev. Lett. {\bf 78}, 310 (1997).
\bibitem{Sawada98} A. Sawada, Z.F. Ezawa, H. Ohno, Y. Horikoshi, Y. Ohno,
S. Kishimoto, F. Matsukura, M. Yasumoto, A. Urayama, Phys. Rev. Lett. {\bf 80},
4534 (1998).
\bibitem{Khrapai00} V.S. Khrapai {\it et al.}, Phys. Rev. Lett. {\bf 84},
725 (2000).  
\end {thebibliography}
\end{document}